\documentclass{article}

\usepackage{fullpage}

\usepackage{graphicx}
\usepackage{amssymb, amsmath}
\usepackage{bm}

\usepackage{tikz} % for schematic graphics
\usetikzlibrary{backgrounds}
\usetikzlibrary{calc}

\definecolor{grey}{rgb}{0.5,0.5,0.5}
\definecolor{lightgrey}{rgb}{0.7,0.7,0.7}
\definecolor{verylightgrey}{rgb}{0.9,0.9,0.9}
\definecolor{darkgrey}{rgb}{0.2,0.2,0.2}

\definecolor{grey5}{rgb}{0.9,0.9,0.9}
\definecolor{grey4}{rgb}{0.833,0.833,0.833}
\definecolor{grey3}{rgb}{0.766,0.766,0.766}
\definecolor{grey2}{rgb}{0.7,0.7,0.7}
\definecolor{grey1}{rgb}{0.633,0.633,0.633}

\title{Modeling temporal hypergraphs}

\author{J\"urgen Lerner$^{1}$\\\texttt{juergen.lerner@uni-konstanz.de}
  \and
  Marian-Gabriel H\^{a}ncean$^{2,3}$\\\texttt{gabriel.hancean@sas.unibuc.ro}
  \and
  Matja{\v z} Perc$^{4,5,6,7}$\\\texttt{matjaz.perc@gmail.com} }

\begin{document}

\maketitle

\noindent $^1$Department of Computer and Information Science, University of Konstanz, Germany

\noindent $^2$Department of Sociology, University of Bucharest, Panduri 90-92, Bucharest, 050663, Romania

\noindent $^3$The Center for Innovation in Medicine, Theodor Pallady Blv. 42J, Bucharest, 032266, Romania

\noindent $^4$Faculty of Natural Sciences and Mathematics, University of Maribor, Koro{\v s}ka cesta 160, 2000 Maribor, Slovenia

\noindent $^5$Community Healthcare Center Dr.\ Adolf Drolc Maribor, Ulica talcev 9, 2000 Maribor, Slovenia

\noindent $^6$Department of Physics, Kyung Hee University, 26 Kyungheedae-ro, Seoul 02447, Republic of Korea

\noindent $^7$University College, Korea University, 145 Anam-ro, Seoul 02841, Republic of Korea

\begin{abstract}
  Networks representing social, biological, technological or other systems are often characterized by higher-order interaction involving any number of nodes. Temporal hypergraphs are given by ordered sequences of hyperedges representing sets of nodes interacting at given points in time. In this paper we discuss how a recently proposed model family for time-stamped hyperedges -- relational hyperevent models (RHEM) -- can be employed to define tailored null distributions for temporal hypergraphs and to test and control for complex dependencies in hypergraph dynamics. RHEM can be specified with a given vector of temporal hyperedge statistics -- functions that quantify the structural position of hyperedges in the history of previous hyperedges -- and equate expected values of these statistics with their empirically observed values. This allows, for instance, to analyze the overrepresentation or underrepresentation of temporal hyperedge configurations in a model that reproduces the observed distributions of possibly complex sub-configurations, including but going beyond node degrees. Concrete examples include, but are not limited to, preferential attachment, repetition of subsets of any given size, triadic closure, homophily, and degree assortativity for subsets of any order.

\end{abstract}

%------------------------------------
\section{Introduction}
\label{sec:intro}

Graphs -- collections of nodes and edges, where each edge connects two nodes -- can represent networks of interconnected entities %and are studied in numerous disciplines %including physics, engineering, computer science, biology, chemistry, and the social sciences
\cite{boccaletti2006complex,wasserman1994social}. Hypergraphs \cite{berge1989hypergraphs} generalize graphs by containing hyperedges that can connect any number of nodes. While having been suggested for network analysis decades ago \cite{seidman1981structures}, hypergraphs received a recent surge of interest \cite{ghoshal2009random,vazquez2009finding,coutinho2020covering,battiston2020networks,sun2021higher,vazquez2023complex}, in part due to the insight that crucial information gets lost when forcing hyperedges into collections of dyadic edges \cite{chodrow2020annotated,lerner2022dynamic}. Examples of complex systems that can be represented by hypergraphs are numerous and include networks of meetings or social events, emails, coauthoring and citation, co-offending, and chemical reactions \cite{newman2001structure,barabasi2002evolution,freeman2003finding,newman2004coauthorship,martin2013coauthorship,jost2019hypergraph,bright2024investigating}.
It has also been shown that higher-order interactions in evolutionary games can have implications on the emergence of collaboration \cite{guo2025evolutionary,wang2025evolutionary}.
Many of these examples give rise to temporal networks \cite{holme2012temporal,masuda2016guide}, in this case, hypergraphs whose hyperedges are time-stamped or time-ordered \cite{lee2023temporal}. 

Quantitative properties of real-world networks are often compared to expected quantities in random graphs that reproduce some observed properties but randomize others. For example, if a network has more closed triangles and a skewed degree distribution compared to uniform random graphs \cite{gilbert1959random,erdos1959random}, then it is informative to analyze whether the degree distribution alone can already explain the higher number of triangles, or whether the latter comprises an additional significant observation beyond what is implied by node degrees. For this purpose one might employ random graphs that reproduce the observed degree sequence (exactly or in expectation) but that randomize everything else \cite{newman2001random} -- a type of random graph models that can also be defined for random hypergraphs with given hyperedge sizes and node degrees (\emph{bipartite configuration model}). Random models, for graphs or for hypergraphs, \cite{newman2001random,ghoshal2009random,chodrow2020configuration,chodrow2020annotated,ruggeri2024framework,huang2024higher} may serve different purposes, which however are interrelated. Frequent use cases employ random graph models (i) as null distributions to compare observed values with expected values, e.\,g., \cite{uzzi2013atypical}, (ii) to find or test network effects, such as preferential attachment, homophily, reciprocity, assortative mixing, or triadic closure, e.\,g., \cite{lusher2013exponential,snijders2005models,bianchi2024relational}, (iii) to generate random (hyper)graphs, e.\,g., \cite{benson2018sequences,kook2020evolution,lee2021hyperedges,ruggeri2024framework}, or (iv) to predict (hyper)graphs, e.\,g., \cite{martinez2016survey,benson2018sequences,yadati2020nhp,chen2023survey}.

In this paper we review a recently proposed family of random models for temporal hypergraphs -- relational hyperevent models (RHEM) \cite{lerner2022dynamic,lerner2023relational} -- and illustrate their use to define tailored null distributions and to test network effects in temporal hypergraphs. RHEM can be specified with a given vector of temporal hyperedge statistics -- functions that quantify how hyperedges are positioned in the history of previous hyperedges -- and successively model the probability of the next hyperedge to reproduce the observed statistics in expectation. We argue that RHEM are a valuable addition to available random models for temporal hypergraphs (see a comprehensive list in \cite{battiston2020networks}) due to the following properties. (i) RHEM are a versatile model family that can be specified with any given vector of temporal hypergraph pattern, including but not limited to preferential attachment, partial repetition of subsets of any order, homophily, assortative mixing, or triadic closure. This allows, for instance, to analyze the overrepresentation or underrepresentation of temporal hyperedge configurations in a model that reproduces the observed distributions of potentially complex sub-configurations. (ii) RHEM can be fitted to real-world networks to equate expected values and observed values in the given vector of temporal hyperedge statistics, as formalized in Eq.~(\ref{eq:expectation}). (iii) RHEM can be generalized in several directions, including directed hypergraphs, multilayer or multimode hypergraphs, and to hypergraphs with typed, labeled, signed, or weighted hyperedges; compare Sect.~\ref{sec:variants}. (iv) From a statistical perspective, RHEM belong to the family of Cox regression models \cite{cox1972regression}, which gives access to a rich set of theoretical results and available software \cite{aalen2008survival,therneau2000modeling,therneau2024survival}.

RHEM generalize (dyadic) relational event models \cite{bianchi2024relational} for time-stamped events having one sender and one receiver to interaction events involving any number of nodes. Specified (only) with subset repetition of various order, RHEM are conceptually similar to the model for ``sequences of sets'' by \cite{benson2018sequences}, which generates the next hyperedge by randomly copying subsets of nodes from previous hyperedges. However, RHEM can be specified with any vector of temporal hyperedge statistics, not just subset repetition. RHEM have been applied to networks in various fields of the social sciences, including the analysis of contact diaries of historical persons, multicast communication (email) networks, coauthoring and citation networks, contact nomination networks of infected persons, criminal networks,  co-offending networks, and artistic production \cite{lerner2021dynamic,lerner2023relational,lerner2023micro,lerner2025relational,hancean2021role,hancean2022occupations,bright2024investigating,bright2024examining,bright2024offence,burgdorf2024communities}. In all these references, the core purpose of RHEM was to test and control for hypothetical network patterns in temporal hypergraphs, while their use to define null distributions has not been discussed. 

For illustration we use as a running example the network of actor co-appearance in chapters of Victor Hugo's novel \emph{Les Mis\'erables}, compiled by Donald Knuth \cite{knuth1993stanford}. While the main motivation for this analysis is to illustrate the model family, it nevertheless gives further empirical support to the insight that when quantifying the overrepresentation or underrepresentation of given hypergraph configurations, such as filled triads, one should not only control for observed node degrees but also for the frequencies of higher-order subsets, such as dyads contained in the respective triad; compare \cite{sarker2024higher}. 

%------------------------------------
\section{\label{sec:prelim}Preliminaries}

%------------------------------------
\subsection{\label{sec:hypergraphs}Hypergraphs and temporal hypergraphs}

A \emph{hypergraph} is a pair $(\mathcal{I},H)$, where $\mathcal{I}=\{i_1,\dots,i_N\}$ is a set of \emph{nodes} (also denoted as \emph{actors} in our illustrating example) and $H\subseteq\mathcal{P}(\mathcal{I})$ is a set of \emph{hyperedges}, that is, subsets of the node set. The \emph{size} of a hyperedge $I\in H$ is the number of its nodes (i.\,e., $size(I)=|I|$) and the \emph{degree} of a node $i\in\mathcal{I}$ in a hypergraph $(\mathcal{I},H)$ is the number of hyperedges containing $i$ (i.\,e., $deg(i)=|\{I\in H\colon i\in I\}|$). More generally, the \emph{degree} of a set of nodes $I\subseteq\mathcal{I}$ is the number of hyperedges jointly containing all nodes in $I$ (i.\,e., $deg(I)=|\{I'\in H\colon I\subseteq I'\}|$).

Many real-world hypergraphs, including meetings, coauthoring, or email communication, represent group interaction occurring at dedicated points in time. A \emph{(relational) hyperevent} is a pair $e=(t,I)$, where $t$ gives the time of the event and $I\subseteq\mathcal{I}$ is the hyperedge containing the nodes participating in the event. A \emph{temporal hypergraph} $(\mathcal{I},E)$ is a set of nodes $\mathcal{I}$ together with a time-ordered sequence of hyperevents
\[
E=[e_1=(t_1,I_1),\dots,e_M=(t_M,I_M)]\enspace,
\]
where, for each $m=1,\dots,M$, it is $I_m\subseteq\mathcal{I}$.
  We note that ``time'' may, but does not have to, represent ``time shown on a clock''. For instance, time may also just be a counter marking the order of arrival of hyperedges. Throughout most of this paper, we assume that time stamps are strictly ordered ($t_1<t_2<\dots<t_M$), that is, there are no simultaneous events. (Note, however, the discussion on ``tied event times'' given in Sect.~\ref{sec:variants}.) We further note that hyperedges may be exactly repeated in $E$, that is, it may be $I_m=I_{m'}$ for different indices $m\neq m'$. 

  For a given sequence of relational hyperevents $E$, that is, a temporal hypergraph, and a point in time $t$, let $E_{<t}=\{(t_m,I_m)\in E\colon t_m<t\}$ denote the \emph{history}, also denoted as \emph{sequence of prior events}, defined as the sub-sequence of all hyperevents that arrive strictly before $t$. We use this notation to define temporally changing degrees of node sets. For a set of nodes $I\subseteq\mathcal{I}$ and a point in time $t$, the \emph{prior degree} of $I$ at $t$ is defined as
\[
deg(I,E_{<t})=|\{(t_m,I_m)\in E_{<t}\colon I\subseteq I_m\}|\enspace,
\]
that is, the number of all hyperevents arriving before $t$ and containing the entire set $I$.

In hypergraphs there is an important distinction between ``closed triads'', sets of three nodes that are pairwise contained in some hyperedges, and ``filled triads'' \cite{sarker2024higher}, sets of three nodes that are jointly members of at least one hyperedge.
%Formally, the \emph{filled triad weight} of a set of three nodes $\{i_1,i_2,i_3\}$ is the degree $deg(\{i_1,i_2,i_3\})$ and the set is called a \emph{filled triad} if this weight is positive. The \emph{closed triad weight} is defined by
%\[
%\min[deg(\{i_1,i_2\}),deg(\{i_1,i_3\}),deg(\{i_2,i_3\})]
%\]
%and the set is called a \emph{closed triad} if this weight is positive.
A filled triad is necessarily a closed triad but not the other way round. If all (hyper)edges have size two, filled triads are not possible but there can still be closed triads.

%------------------------------------
\subsection{\label{sec:data_lesmis}Data: Les Mis\'erables}

\begin{figure}
  \centering
  \includegraphics[width=0.5\linewidth]{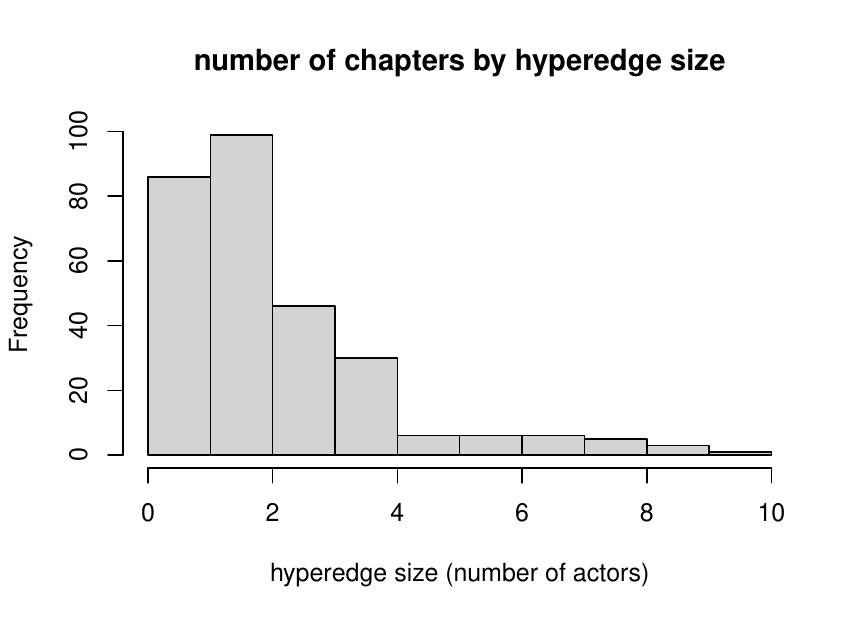}%
  \caption{\label{fig:hyperedge_sizes}Number of chapters with a given number of actors (i.\,e., given the hyperedge size).}
\end{figure}

For illustration we use as a running example the network of actor co-appearance in chapters of Victor Hugo's novel \emph{Les Mis\'erables}, compiled by Donald Knuth \cite{knuth1993stanford}. The data comprises 80 actors co-appearing in one or several of 288 chapters. Each of these chapters defines one hyperevent $e=(t,I)$, whose hyperedge $I$ contains the actors starring in that chapter and where the time stamps are counters for chapters $t\in\{1,\dots,288\}$ in the order in which they appear in the novel. The number of actors per chapter, that is, the hyperedge size varies from one to ten, where chapters with more than four actors are rare; see Fig.~\ref{fig:hyperedge_sizes}.

%---------------------------------------------------
%---------------------------------------------------

To illustrate the use of exogenous node attributes, we augment Knuth's \emph{Les Mis\'erables} data by a binary gender variable, $female(i)$, which is one if and only if actor $i$ is female. We believe that, using actors' names, Knuth's short description, and our own reading of the novel, we can derive this variable with good reliability. (We note that our analysis is given for illustrating the model and not for making any claims about the demography of the novel's characters.) Among the 80 actors, there are 28 female (35\%). For reproducibility, we provide the preprocessed data, augmented by the $female$ variable, in the file \texttt{jean\_events.csv}, linked from the \texttt{eventnet} Webpage \footnote{\texttt{https://github.com/juergenlerner/eventnet/tree/master/data/les\_miserables}}.

The \emph{Les Mis\'erables} data is convenient for illustrating models in this paper, since it is established in the networks literature, e.\,g., \cite{newman2004finding,min2019modeling}, and since the actors and their relations are part of a story, allowing to anecdotally interpret certain findings. We list applications of RHEM to several other real-world temporal hypergraphs in Sect.~\ref{sec:variants}. 

%book read: \cite{hugo1966miserables1,hugo1966miserables2}

%------------------------------------
\section{\label{sec:rhem}RHEM for temporal hypergraphs}

%------------------------------------
\subsection{\label{sec:framework}Model framework}

Models for temporal hypergraphs often successively specify probabilities of the ``next'' hyperedge, given the history of all previous hyperedges. RHEM specify these conditional probabilities dependent on a vector of temporal hyperedge statistics
\[
\bm{x}(\cdot,E_{<t_m})\colon\mathcal{P}(\mathcal{I})\to\mathbb{R}^k\,;\;
I\mapsto\bm{x}(I,E_{<t_m})
\]
that quantify selected properties of any possible next hyperedge $I\subseteq\mathcal{I}$ in relation to the respective history $E_{<t_m}$. Examples for such statistics are the sum of the prior node degrees $deg(i,E_{<t})$ for all $i\in I$, the sum of the prior degrees $deg(\{i_1,i_2\},E_{<t})$ for all pairs of nodes $\{i_1,i_2\}\in\binom{I}{2}$, sums of the degrees of larger subsets, measures of triadic closure, degree assortativity, and homophily.  

To illustrate the use of such statistics, and their connection to hypothesized patterns in temporal hypergraph, we take a one-dimensional $\bm{x}(I,E_{<t_m})$, defined to be the sum of prior degrees of members of $I$:
\begin{equation}
  \label{eq:degsum}
  \bm{x}(I,E_{<t_m})=degsum(I,E_{<t})=\sum_{i\in I}deg(i,E_{<t})
\end{equation}
and illustrate the connection of that statistic to a hypothetical preferential attachment effect. Under a uniform null model $M_0$ that selects members of the next hyperedge uniformly at random, and if we fix the size $|I|$ of the next hyperedge, the expected value would be equal to $|I|$ times the average prior node degree at $t$:
\[
\mathbb{E}_{M_0}\left[degsum(\cdot,E_{<t})\right]=|I|\sum_{I_m\in E_{<t}}\frac{|I_m|}{|\mathcal{I}|}\enspace.
\]
If, however, we assume a preferential attachment mechanism in which nodes with higher prior degrees are selected into the next hyperedge with higher probability, then the observed value of $degsum$ would likely be higher than the expected value under the uniform null model, providing empirical evidence for preferential attachment of individual nodes. The RHEM family, defined below, can be used to fit a model $M_1$ such that the expected value of $degsum$ over the entire sequence of events under $M_1$ is equal to the observed value and this model, in turn, can be used to test whether preferential attachment of individual nodes is already sufficient to explain other patterns in temporal hypergraphs, such as preferential attachment of higher-order sets, assortativity, or triadic closure. In the following we review the definition of RHEM and illustrate its use as a general model family for temporal hypergraphs designed to test hypothetical patterns, simultaneously controlling for a given vector of temporal hyperedge statistics $\bm{x}(\cdot,E_{<t_m})$.

Concretely, RHEM \cite{lerner2021dynamic} specify the probability of a time-ordered sequence of hyperevents $E$, dependent on a vector of statistics $\bm{x}(\cdot,E_{<t_m})$ and an associated parameter vector $\bm{\beta}$, by
\begin{equation}\label{eq:prob_sequence}
P(E,\bm{\beta})=\prod_{m=1}^MP(I_m|E_{<t_m},\bm{\beta}),
\end{equation}
that is, factorizing the probability of the entire sequence into the product of conditional probabilities of the next hyperedge, given the history of prior events.
The conditional probability of the next hyperedge $I_m$, in turn, is specified by
\begin{equation}\label{eq:prob_next}
P(I_m|E_{<t_m},\bm{\beta})=\frac{\exp[\bm{\beta}\cdot\bm{x}(I_m,E_{<t_m})]}
{\sum_{I\in\binom{\mathcal{I}}{|I_m|}}\exp[\bm{\beta}\cdot\bm{x}(I,E_{<t_m})]}
\enspace.
\end{equation}
Equation~(\ref{eq:prob_next}) is one of the most common specifications of discrete choice models \cite{agresti2015foundations}, where an instance $I_m$ is selected out of a space of possible instances $\binom{\mathcal{I}}{|I_m|}$, that is, all possible subsets of nodes of the same size as $I_m$. We note that the conditioning on observed hyperedge sizes mirrors the common approach of many models for static or temporal hypergraphs, such as the bipartite configuration model \cite{newman2001random}.

Conveniently, the specification in Eqs.~(\ref{eq:prob_sequence}) and~(\ref{eq:prob_next}) is identical to common specifications of the partial likelihood of Cox proportional hazard (CoxPH) models \cite{cox1972regression,aalen2008survival,lerner2023relational}, one of the most established models for event history analysis. This gives the exponential of the parameters, $\exp(\beta_h)$, the interpretation as ``hazard ratios'', that is, factors by which the event rate on a set of nodes $I$ is expected to increase (or decrease), if the associated hyperedge statistic $x_h(I,E_{<t})$ increases by one unit. The connection to the CoxPH model also implies an abundance of existing software packages to estimate parameters and a wealth of theoretical results. Of special importance for RHEM are results about the consistency of parameters estimated from sampled likelihoods -- obtained by case-control sampling \cite{borgan1995methods,lerner2023relational} --, where the index set $\binom{\mathcal{I}}{|I_m|}$ in the denominator of (\ref{eq:prob_next}) gets replaced by a sampled set containing the hyperedge $I_m$ of the observed event (the ``case'') plus $K$ non-event hyperedges (``controls'') $\{I_1,\dots,I_K\}$ which are sampled uniformly at random from $\binom{\mathcal{I}}{|I_m|}$. Estimation via sampled likelihoods is necessary since the entire set $\binom{\mathcal{I}}{|I_m|}$ is often prohibitively large for all but the smallest node sets $\mathcal{I}$ and hyperedge sizes $|I_m|$. Using case-control sampling, RHEM have been applied to networks containing more than a million of nodes and events, e.\,g., \cite{lerner2025relational}.

Exponential-family models, like the one specified in Eqs.~(\ref{eq:prob_sequence}) and~(\ref{eq:prob_next}), have the property that the observed explanatory variables are equal to their expectation in the model with parameters estimated by maximizing the likelihood. (This well-known fact can be derived by setting the partial derivatives of the log-likelihood to zero.) Concretely, for the maximum likelihood estimates $\hat{\bm{\beta}}$ it is
\begin{equation}\label{eq:expectation}
\sum_{m=1}^M\bm{x}(I_m,E_{<t_m})=
\sum_{m=1}^M\mathbb{E}_{\hat{\bm{\beta}}}\left[\bm{x}(\cdot,E_{<t_m})\right]\enspace.
\end{equation}
This is a useful result as it enables us to find parameters such that the model reproduces in expectation any given vector of temporal hyperedge statistics. Such a model can be used as a null model matching selected patterns of observed temporal hypergraphs -- and then checking whether further observed properties are already explained by that model, or whether they deviate from what would be expected under a (non-uniform and potentially complex) null model.

%------------------------------------
\subsection{\label{sec:statistics} Commonly used temporal hyperedge statistics}

In the following, we introduce examples of frequently used temporal hyperedge statistics, building increasingly complex models, and briefly discuss selected findings on the \emph{Les Mis\'erables} data for illustration. We estimate model parameters by maximum likelihood estimation with case-control sampling, where we bound the number of selected non-events per event to $\binom{80}{3}=82,160$. This means that for events $(t_m,I_m)$ of size one, two, and three we take the entire set of subsets $\binom{\mathcal{I}}{|I_m|}$ -- compare the index set in the denominator of (\ref{eq:prob_next}) -- and for larger events we bound the set of sampled non-events to the number of different triads of nodes. Sampling non-events and computing hyperedge statistics is done with the \texttt{eventnet} software \cite{lerner2023relational}. Given the table of hyperedge statistics, parameter estimation is done with the \texttt{coxph} function in the \texttt{R} package \texttt{survival} \cite{therneau2000modeling,therneau2024survival}.

\paragraph*{Repetition of subsets of various order.}
As it is common in real-world hypergraphs, node degrees are skewedly distributed in the \emph{Les Mis\'erables} data. For example, the most frequent actors are three of the novel's main characters: \emph{Jean Valjean} (JV, appearing in 113 chapters), \emph{Marius} (MA, 77 chapters), and \emph{Cosette} (CO, 55 chapters), largely exceeding the  mean number of $9.1$ chapters per actor.

Models for static (time-independent) hypergraphs, like the bipartite configuration model, often constrain the probability space to reproduce the observed degree sequence exactly or in expectation. In the logic of temporal hypergraph models we can obtain a skewed degree distribution by a preferential attachment mechanism, letting nodes with a larger number of past events be selected as participants of the next events with a higher probability. This can be achieved by a temporal hyperedge statistic $x(I,E_{<t})$ counting the past events of the nodes in $I$. Since we also consider preferential attachment for higher-order subsets (e.\,g., pairs and triples), we introduce \emph{subset repetition of order $k$}, for varying $k\geq 1$, by
\begin{equation}
  \label{eq:subrep}
  subrep^{(k)}(I,E_{<t})=\sum_{I'\in\binom{I}{k}}deg(I',E_{<t})\enspace.
\end{equation}
(The index set $\binom{I}{k}$ is the set of all subsets of size $k$, that is, $\binom{I}{k}=\{I'\subseteq I\colon |I'|=k\}$.)
We point out that the previously discussed $degsum$ statistic is identical to $subrep^{(1)}$.
In general, subset repetition of order $k$ on a set of nodes $I$ is the sum of the prior degrees over all $k$-element subsets of $I$.
Figure~\ref{fig:illustration} illustrates subset repetition of order one, two, and three. For example, subset repetition of order one takes higher values on the hyperedge $I'=\{D,E,F\}$, subset repetition of order three is higher for $I=\{A,B,C\}$, and both hyperedges have the identical value for subset repetition of order two.

%---------------------------------------------------
\begin{figure}
  \begin{center}
    \begin{tikzpicture}
      \node at (0,0) (list) {\footnotesize$\begin{array}{l}
          e_1=(t_1,\{A,B,C\})\\
          e_2=(t_2,\{D,E\})\\
          e_3=(t_3,\{D,F\})\\
          e_4=(t_4,\{E,F\})\\
        \end{array}$};
      \node at (4.5,0) (hyper) {\footnotesize
        \begin{tikzpicture}[scale=0.7]
          \tikzstyle{actor}=[circle,minimum size=1mm];
          \node[actor] at (-3.5,1) (a1) {$A$};
          \node[actor] at (-3.5,-1) (b1) {$B$};
          \node[actor] at (-1.5,0) (c1) {$C$};
          \node[actor] at (2,1) (a2) {$E$};
          \node[actor] at (2,-1) (b2) {$F$};
          \node[actor] at (0,0) (c2) {$D$};

          \draw[color=darkgrey,line width=1pt,dashed] (a1) to (c1);
          \draw[color=darkgrey,line width=1pt,dashed] (a1) to (b1);
          \draw[color=darkgrey,line width=1pt,dashed] (b1) to (c1);

          \draw[color=darkgrey,line width=1pt,dashed] (a2) to (c2);
          \draw[color=darkgrey,line width=1pt,dashed] (a2) to (b2);
          \draw[color=darkgrey,line width=1pt,dashed] (b2) to (c2);

          \begin{pgfonlayer}{background}
            \foreach \nodename in {a1,b1,c1,a2,b2,c2} {
              \coordinate (\nodename') at (\nodename);
            }
            \path[fill=grey1,draw=grey1,line width=0.7cm, line cap=round, line join=round] 
            (a1') to (b1') to (c1') to (a1') -- cycle;
            \path[fill=grey2,draw=grey2,line width=0.55cm, line cap=round, line join=round] 
            (a2') to (c2') to (a2') -- cycle;
            \path[fill=grey3,draw=grey3,line width=0.55cm, line cap=round, line join=round] 
            (b2') to (c2') to (b2') -- cycle;
            \path[fill=grey4,draw=grey4,line width=0.45cm, line cap=round, line join=round] 
            (a2') to (b2') to (a2') -- cycle;
          \end{pgfonlayer}
        \end{tikzpicture}
      };
    \end{tikzpicture}
  \end{center}
  \caption{\label{fig:illustration} (Adapted from \cite{lerner2022dynamic}.) Illustrating subset repetition based on four prior events $e_1,\dots,e_4$ on the hyperedges $I=\{A,B,C\}$ and $I'=\{D,E,F\}$ at time $t>t_4$. For $I$ it is $subrep^{(1)}(I',E_{<t})=3$, $subrep^{(2)}(I',E_{<t})=3$, and $subrep^{(3)}(I',E_{<t})=1$. For $I'$ it is $subrep^{(1)}(I',E_{<t})=6$, $subrep^{(2)}(I',E_{<t})=3$, and $subrep^{(3)}(I',E_{<t})=1$. Moreover, Event~$e_4$ is an illustration of triadic closure (defined further below), as the event connecting Nodes $E$ and $F$, closes a two-path $E$--$D$--$F$, which has been established in two previous events, $e_2$ and $e_3$.}
\end{figure}
%---------------------------------------------------

Models reported in Table~\ref{tab:models_subrep} reveal that the \emph{Les Mis\'erables} data provides evidence for preferential attachment of subsets containing one, two, and three actors. Individual actors with a higher number of past events are selected into future events with a higher probability. Controlling for this individual popularity effect, pairs of actors that co-appear in more previous chapters are jointly selected into future chapters with a higher probability. Controlling for individual and pairwise popularity effects, triples of actors that co-appear in more previous chapters tend to co-appear in future chapters with a higher probability. 

%---------------------------------------------------
\begin{table}[ht]
  \caption{\label{tab:models_subrep} Parameters of RHEM estimated from the \emph{Les Mis\'erables} data. All models have been estimated with 288 events (chapters) and a cumulative number of $8,782,257$ selected non-events.}
  \centering
  \begin{tabular}{l c c c}
    \hline
    & Model 1 & Model 2 & Model 3 \\
    \hline
    subrep$^{(1)}$   & $0.059 \; (0.002)^{***}$ & $0.035 \; (0.003)^{***}$ & $0.048 \; (0.003)^{***}$ \\
    subrep$^{(2)}$   &                          & $0.164 \; (0.010)^{***}$ & $0.096 \; (0.011)^{***}$ \\
    subrep$^{(3)}$   &                          &                          & $0.218 \; (0.024)^{***}$ \\
    \hline
    AIC         & $3935.734$               & $3626.535$               & $3493.831$               \\
    \hline
    \multicolumn{4}{l}{\scriptsize{$^{***}p<0.001$; $^{**}p<0.01$; $^{*}p<0.05$}}
  \end{tabular}
\end{table}
%---------------------------------------------------

Theoretically, we may increase the order of the subset repetition statistic up to the maximum size of hyperedges (beyond which the statistic is necessarily zero). However, in practice estimation becomes unstable, or does not converge at all, if interaction among subsets of size $k$ is too sparse in the given data and if subset repetition of order $k$ or higher is included in the model (see the discussion of this aspect given in \cite{lerner2023relational}). When we fit models with subset repetition of higher and higher order to the \emph{Les Mis\'erables} data, we find that already subset repetition of order four is non-significant and starting from subset repetition of order six, the estimation does not converge. We point out that \cite{lerner2023micro} estimated subset repetition up to order ten in a coauthoring network in which hyperedge sizes (i.\,e., number of authors per paper) up to 20 or beyond are rather frequent.

We can use any estimated RHEM also to compute the expected values of specifically chosen sets of nodes. For instance, we may use Model~1 in Table~\ref{tab:models_subrep}, specified only with preferential attachment of individual nodes, to check whether specific pairs of actors co-appear in chapters more or less frequently than suggested by their individual frequencies. We find that the most overrepresented dyad is JV-CO (expected to co-appear in 13 chapters vs.\ an observed number of 33 co-appearances), followed by EN-CR (expected 0.33; observed 18), EN-CM (expected 0.30, observed 15), and MA-CO (expected 4.7, observed 19). The dyad JV-MA, which is the one with the highest expected number of co-appearances (20 expected vs.\ 18 observed) is only on position 3,125 (among 3,160 dyads) when ranking dyads by their overrepresentation. So JV-MA is among the most underrepresented dyads. By Heider's balance theory \cite{heider1946attitudes,cartwright1956structural,lerner2016structural,kirkley2019balance}, and labeling dyads with the differences of actually observed co-appearances minus the expected number, the triangle JV-MA-CO is an imbalanced triangle, where \emph{Jean Valjean} and \emph{Marius} are both positively linked to \emph{Cosette}, but meet themselves less often than expected, given their individual frequencies.

We can also standardize differences between observed and expected values, dividing by the estimated standard deviations, to obtain a \emph{z-score} measure. For example, we get a z-score of 5.4 for the JV-CO dyad and a z-score of 6.7 for MA-CO, so that both dyads are significantly overrepresented. On the other hand, the z-score of JV-MA is $-0.45$, so that this dyad is not significantly underrepresented but the difference to the expected value could be explained by random chance.

Turning to overrepresentation or underrepresentation of filled triads, according to Model~1, we find that the most overrepresented triad is among three of the ``friends of the ABC'', namely EN-CM-CR, which co-appear in 12 chapters (the maximum value over all triads) and are expected to co-appear in only 0.01 chapters. The triad formed by the three most frequent actors JV-MA-CO is expected to appear in 3.3 chapters but does appear in 6 chapters in the actual novel, placing this triad on position 164 (among 82,160 triads) when ranking triads by their overrepresentation according to Model~1. The assessment of the JV-MA-CO triad changes remarkably if we additionally control for a preferential attachment effect of actor pairs. According to Model~2 in Table~\ref{tab:models_subrep}, specified with subset repetition of order one and two, JV-MA-CO is the most underrepresented triad, with an expected number of 16 co-appearances but only 6 co-appearances in the actually observed data. This difference to the expectation of Model~1 can be explained by the finding that two of the three contained dyads, namely JV-CO and MA-CO, are strongly overrepresented (controlling for individual frequencies). Considering also previous dyadic co-appearances, the probability that all three, JV, MA, and CO, jointly appear in a chapter increases. This illustrates that if we want to assess the relative frequency of higher-order structures (such as filled triads), we should not only control for the prior degree of individual nodes but also for the prior occurrence of more complex sub-structures, such as dyads that are subsets of the respective triads. We can find a similar reasoning in \cite{sarker2024higher} who argue that to test higher-order homophily in hypergraphs one has to control for lower-order homophily.

%------------------------------------
\paragraph*{Triadic closure.}
Triadic closure is the effect that two nodes $i_1$ and $i_2$ both had a prior contact with a common third node $i_3$, possibly in different past events, and then $i_1$ and $i_2$ meet in a future event. Triadic closure is a different effect than the repetition of filled triads. Indeed, even if all (hyper)edges have size two, there may still be triadic closure, but there cannot be any filled triads. We can test with RHEM whether a temporal hypergraph provides evidence for triadic closure, on top of what is explained by preferential attachment of individual nodes, dyads, triads, or any other size of higher-order subsets. For such tests we include a temporal hyperedge statistic defined by:
\begin{equation}\label{eq:closure}
   closure(I,E_{<t})=\sum_{\{i_1,i_2\}\in\binom{I}{2} \atop i_3\in\mathcal{I}\colon i_3\neq i_1,i_2}
  \min[deg(\{i_1,i_3\},E_{<t}),deg(\{i_2,i_3\},E_{<t})]\enspace.
\end{equation}
For each dyad $\{i_1,i_2\}$ in $I$ we look for ``third nodes'' $i_3$ (not necessarily in $I$) and then we weight the two-path $i_1$--$i_3$--$i_2$ by the minimum prior degree of $\{i_1,i_3\}$ and $\{i_2,i_3\}$. We then add up these weights over all dyads included in $I$ and all third nodes. Closure is illustrated in the stylized example shown in Fig.~\ref{fig:illustration}.

Models in Table~\ref{tab:models_other} reveal that closure events are overrepresented in the \emph{Les Mis\'erables} data. We can also find specific chapters that are particularly well explained by triadic closure. Among the strongest is Chapter~4.15.2 in which \emph{Jean Valjean} (JV) and \emph{Gavroche} (GA) meet for the first time. Before that, both had prior encounters with various other common third actors, albeit in different chapters. The strongest ``bundle'' of prior two-paths connecting JV and GA is via \emph{Th\'enardier/Jondrette} (TH/JD) and the gang of criminals (including MO, BB, BJ, GU) he collaborates with to extort money from JV in Chapter~3.8.20. \textit{Gavroche}, on the other hand, assists in helping his (at the end unthankful) father TH to escape from prison in Chapter~4.6.3, together with MO, BB, BJ, GU, creating several two-paths between GA and JV before their encounter in Chapter~4.15.2.

%---------------------------------------------------
\begin{table}[ht]
  \caption{\label{tab:models_other} Parameters of RHEM estimated from the \emph{Les Mis\'erables} data. All models have been estimated with 288 events (chapters) and a cumulative number of $8,782,257$ selected non-events. }
  \centering
\begin{tabular}{l c c c}
\hline
 & Model 4 & Model 5 & Model 6 \\
\hline
subrep$^{(1)}$           & $0.048 \; (0.003)^{***}$ & $0.055 \; (0.003)^{***}$  & $0.054 \; (0.003)^{***}$  \\
subrep$^{(2)}$           & $0.083 \; (0.013)^{***}$ & $0.102 \; (0.013)^{***}$  & $0.113 \; (0.014)^{***}$  \\
subrep$^{(3)}$           & $0.126 \; (0.032)^{***}$ & $0.116 \; (0.035)^{***}$  & $0.121 \; (0.035)^{***}$  \\
closure             & $0.008 \; (0.003)^{**}$  & $0.008 \; (0.003)^{**}$   & $0.006 \; (0.003)^{*}$    \\
assort$^{(1)}$ &                          & $0.010 \; (0.002)^{***}$ & $0.010 \; (0.002)^{***}$ \\
assort$^{(2)}$ &                          & $0.006 \; (0.002)^{***}$ & $0.005 \; (0.002)^{**}$  \\
female              &                          &                           & $-0.118 \; (0.112)$       \\
homophily           &                          &                           & $0.183 \; (0.068)^{**}$  \\
\hline
AIC                 & $3495.341$               & $3432.036$                & $3420.067$                \\
\hline
\multicolumn{4}{l}{\scriptsize{$^{***}p<0.001$; $^{**}p<0.01$; $^{*}p<0.05$}}
\end{tabular}
\end{table}
%---------------------------------------------------

%------------------------------------
\paragraph*{Assortativity of various order.}
Degree assortativity is the pattern that nodes tend to connect with other nodes having a similar degree. As it is the case for preferential attachment (implemented in RHEM via subset repetition), assortativity in temporal hypergraphs can be defined for subsets of any size, not just for individual nodes. We define a temporal hyperedge statistic \emph{assortativity of order $k$}, for varying $k\geq 1$, by
%\begin{equation}
%  \label{eq:assort}
%  assort^{(k)}(I,E_{<t})=-\sum_{\{I'\neq I''\}\subseteq\binom{I}{k}}|deg(I')-deg(I'')|\enspace,
%\end{equation}
\begin{equation}
  \label{eq:assort}
  assort^{(k)}(I,E_{<t})=
  -\sum_{
    \{I',I''\}\in\left(
    \genfrac{}{}{0pt}{0}{\binom{I}{k}}{2}
    \right)
  }
  |deg(I',E_{<t})-deg(I'',E_{<t})|\enspace.
\end{equation}
The summation index runs over all unordered pairs $\{I',I''\}$ of different subsets of size $k$ of the given hyperedge $I$. For each of these pairs we add the absolute difference between the degrees $deg(I',E_{<t})$ and $deg(I'',E_{<t})$. The resulting sum would be a measure of disassortativity as it gets larger when the prior degrees of the contained subsets are more different. We therefore take the negative sum to get a measure of degree assortativity. It is not necessary to add a constant in order to make the statistic positive, since any additive constant would cancel out in (\ref{eq:prob_next}). We note that $assort^{(k)}(I,E_{<t})$ can assume non-zero values only if $|I|\geq k+1$, since there have to be more than one subset of size $k$ to get any pairs of such subsets.

In Model~5 in Table~\ref{tab:models_other} we find that there is assortativity of individual nodes, so that actors tend to co-appear together with other actors having a similar prior degree (i.\,e., that appeared in a similar number of previous chapters). On top of that we find degree assortativity for pairs of nodes. This means that dyads tend to co-appear with other dyads having a similar number of previous co-appearances, controlling for assortativity of individual nodes.

%------------------------------------
\paragraph*{Exogenous variables.}
We can define statistics to assess first-order effects and homophily with respect to exogenous variables, such as age or gender. The definition of such statistics is very similar to subset repetition and assortativity of order one. The only change in the definition is that we replace the prior degree by a function giving the values of the respective variable. For example, if we have a binary variable $female(i)$, taking the value one if and only if $i$ is female, we can assess the first-order effect (testing whether females participate in events at a higher rate) via the statistic counting the number of females in a set of nodes $I$ (which we also denote as $female$ by a slight abuse of notation):
\begin{equation}
  \label{eq:female}
  female(I)=\sum_{i\in I}female(i)\enspace.
\end{equation}
The homophily with respect to gender (testing whether females tend to co-participate in events together with other females and/or males tend to co-appear with other males) can be measured with the statistic:
\begin{equation}
  \label{eq:female_homophily}
  homophily(I)=-\sum_{\{i,i'\}\in\binom{I}{2}}|female(i)-female(i')|\enspace,
\end{equation}
counting the number of pairs in $I$ that have different gender.

In the \emph{Les Mis\'erables} data (augmented by the $gender$ variable as described above) we find that females are neither significantly overrepresented nor underrepresented in events (compared to their prevalence in the set of all 80 characters), but that there is homophily with respect to the ``female'' variable; see Model~6 in Table~\ref{tab:models_other}. This means that actors typically co-appear with other actors of the same gender, more than what would be expected by chance alone.

%------------------------------------
\subsection{\label{sec:event_probs} Event probabilities}

\begin{figure}
  \centering
  \includegraphics[width=0.6\linewidth]{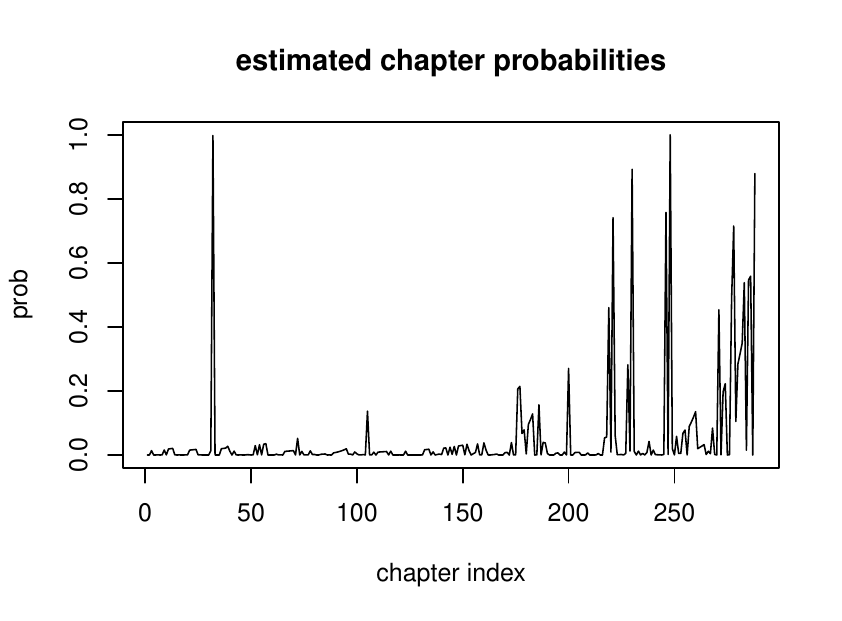}%
  \caption{\label{fig:chapter_probs} Chapter probabilities estimated by Model~3, in Table~\ref{tab:models_subrep}.}
\end{figure}

RHEM can also be used to assess the estimated probabilities of observed hyperedges by computing values via Eq.~(\ref{eq:prob_next}). We can interpret these probabilities at face value or relative to the probabilities of sampled alternative node sets that could have constituted the next hyperedge (``non-events''). Figure~\ref{fig:chapter_probs} displays the probabilities of the 288 chapters in \emph{Les Mis\'erables} estimated by Model~3, in Table~\ref{tab:models_subrep}. 

%---------------------------------------------------
%---------------------------------------------------

The mean estimated probability over all actual chapters is $5\%$, exceeding even the probability of random guessing of the one-person chapters (that is $1/80$). Ten of the 288 chapters have a probability larger than or equal to $0.5$. Figure~\ref{fig:chapter_probs} shows some chapters of remarkably high probabilities, especially towards the end of the novel, where the model has a longer history to build upon. The two chapters with the highest probabilities ($p>0.99$) are Chapter~1.3.8, featuring the repeated appearance of eight actors (FN, DA, FV, and ZE together with their unfaithful lovers FT, LI, BL, and FA) and Chapter~5.1.21, featuring the re-appearance of six ``friends of the ABC'' together with \textit{Marius}. Another remarkable chapter, regarding estimated probabilities, is the last one (Chapter~5.9.5, $p>0.87$, in Position~4 of 288 when ranking chapters by their predicted probabilities) in which \emph{Jean Valjean}, \emph{Marius}, and \emph{Cosette} are happily reunited, further ``filling'' the triad composed of the three most prominent actors.

\begin{figure}
  \centering
  \includegraphics[width=0.6\linewidth]{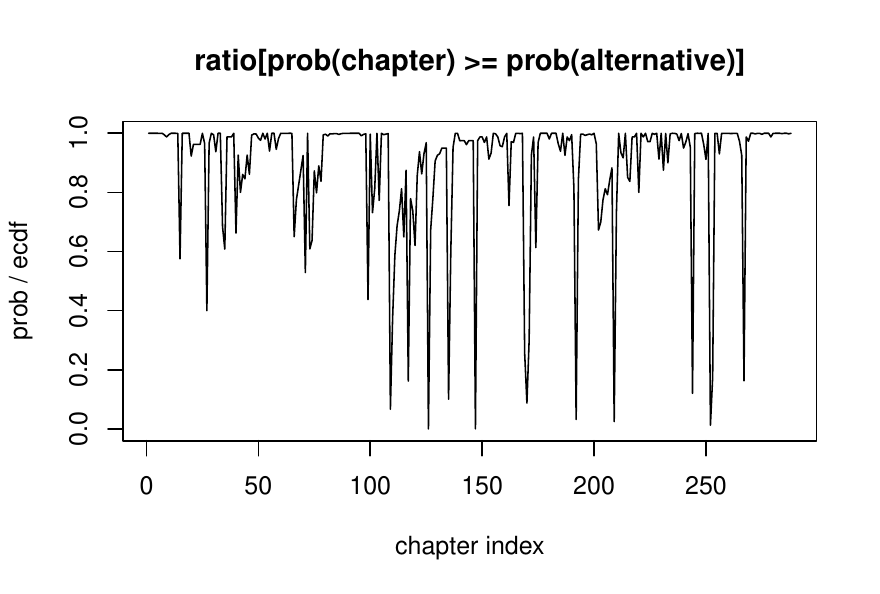}%
  \caption{\label{fig:chapter_ecdf} Probability that the chapter's probability estimated by Model~3, in Table~\ref{tab:models_subrep} is larger than the estimated probability of a sampled alternative ``non-event'' hyperedge of the same size.}
\end{figure}

The probabilities, displayed in Fig.~\ref{fig:chapter_probs}, can be misleading, due to the high number of alternatives, and are also affected by hyperedge size. For example, uniform random guessing of the participant of a one-person chapter yields probability $1/80$, while uniformly guessing the participants of a three-person chapter yields a probability of $\binom{80}{3}^{-1}$, that is, a chance of 1 to 82,160. As an alternative assessment of chapter probability we display in Fig.~\ref{fig:chapter_ecdf} for each chapter the ratio of sampled alternative subsets with an estimated probability lower than or equal to the estimated probability of the hyperedge of the actual chapter. (This is the value of the empirical cumulative distribution function, ECDF, of the actual chapter's probability in the distribution defined by the probabilities of the sampled alternatives.)

Figure~\ref{fig:chapter_ecdf} shows that the predicted probabilities of most of the actual chapters exceed the predicted probabilities of the majority of sampled alternative (mean ECDF is 0.89). 50 chapters are assigned the highest predicted probabilities among all sampled alternatives (ECDF equal 1.0) and 271, that is, all but 17 chapters, have predicted probabilities larger than half of the sampled alternatives (ECDF larger than or equal to 0.5). However, we can also see that throughout the novel there are chapters with low relative predicted probabilities. These are typically chapters introducing a new set of actors, such as Chapter~3.4.1 (ECDF lower than $10^{-4}$ and probability lower than $10^{-16}$) in which nine ``friends of the ABC'' appear for the first time. It is fairly obvious that models relying only on prior chapter compositions cannot forecast the first appearance of actors and even less so the simultaneous appearance of nine previously unseen actors.

%---------------------------------------------------
%---------------------------------------------------

%------------------------------------
\section{\label{sec:variants}Model variants and other data}

The basic RHEM model family for time-stamped hyperedges $(t,I)$, introduced in the previous section, has been adapted and extended in various directions, which we briefly review in the following.

\paragraph*{Directed hyperedges.}
Hyperevents $(t,I)$ modeled so far involve a single hyperedge $I\subseteq\mathcal{I}$ whose members are unordered. This structure applies to hyperevents like meetings \cite{lerner2021dynamic}, co-attendance at social events \cite{lerner2022dynamic}, or team work such as coauthoring \cite{lerner2023micro} or movie production \cite{burgdorf2024communities}. However, there are other types of higher-order networks containing hyperedges that are partitioned into senders and receivers. For example, multicast communication, e.\,g., email, produces communication events of the form $(t,i,J)$, where sender $i\in\mathcal{I}$ sends a message to the set of receivers $J\subseteq\mathcal{I}$ at time $t$. RHEM for sequences of directed multicast events $(t_1,i_1,J_1),\dots,(t_M,i_M,J_M)$ have been proposed in \cite{lerner2023relational}. While the dynamics of the receiver hyperedges $J_m,\,m=1,\dots,M$ may be modeled as it has been discussed above, the direction -- or, from another point of view, the information about the sender $i_m$ -- gives rise to additional hypothetical patterns, such as reciprocation or different variants of triadic closure, including transitive closure and cyclic closure, which are not possible in ``undirected'' hyperevents; see \cite{lerner2023relational} for details. A different application setting in which RHEM for directed hyperevents have been applied are disease-spreading networks representing interaction between an infected person and her recent contacts \cite{hancean2021role,hancean2022occupations}. It would also be possible to develop RHEM for directed hyperevents $(t,I,J)$ having several senders $I\subseteq\mathcal{I}$ and several receivers $J\subseteq\mathcal{I}$ -- for example, the reactants and products of chemical reactions; see the discussion on multimode hyperevents below.

\paragraph*{Multilayer or multimode hyperevents.}
Hypergraphs may be defined on node sets that are partitioned into different types of nodes, often denoted as multilayer or multimode networks \cite{kivela2014multilayer,boccaletti2014structure}. For example, \cite{bright2024examining,bright2024offence} propose models for co-offending events $(t,I,C)$ where $I\subseteq\mathcal{I}$ is the set of actors who, at time $t$, jointly commit a crime that is labeled by a set of crime categories $C\in\mathcal{C}$, taken from a universe of possible crime categories $\mathcal{C}$. The structure of these ``typed events'' resembles directed hyperevents discussed above since each event involves two sets of nodes (or two hyperedges). However, the difference is that in directed hyperevents representing communication events the sender(s) are taken from the same set of nodes $\mathcal{I}$ as the receivers, while in the case of typed events, the participating actors and the labels (e.\,g., crime categories) are selected from disjoint sets. Typed hyperevents give rise to further possible patterns in temporal hypergraphs. For example, multilayer subset repetition statistics can be used to test whether the same (set of) actors repeatedly tend to engage in the same (set of) crime categories. As another example, ``mixed-closure'' statistics can be used to test whether actors ``learn'' to (or are influenced to) engage in new crime categories by their former co-offending partners. The respective statistic is similar to the closure statistic defined in (\ref{eq:closure}), but considers two paths of the kind $i_1$--$i_2$--$c$ indicating that $i_1$ and $i_2$ have co-offended before and that $i_2$ has engaged in crime category $c$ before (possibly in a different event) and then tests whether $i_1$ is more likely to also engage in crime category $c$.

As another example for multilayer hyperevents, \cite{lerner2025relational} propose RHEM for the coevolution of coauthoring and citation networks, where a ``publication event'' $(t,j,I,J)$ indicates that the set of authors $I\subseteq\mathcal{I}$ publish paper $j\in\mathcal{J}$ at time $t$, citing the set of papers $J\subseteq\mathcal{J}$ in its references. (This coding is remarkably similar to the example of complex hypergraphs representing systems of scientific publications, given in Fig.~1 of \cite{vazquez2023complex}.) These multilayer hyperevents for scientific publications allow to additionally test fairly complex mixed-mode patterns, such as whether scientists have a tendency to start coauthoring with those who cited their own prior work or whether they tend to cite the papers of those who cited their own work. It would be possible to further extend such complex publication events, for example to $(t,j,I,J,K,\dots)$, where $K\subseteq\mathcal{K}$ is the set of keywords of the published paper $j$, and so on.

\paragraph*{Weighted or signed hyperevents.}
Hyperevents may also be labeled with numerical information, representing event weights or event signs. For example, \cite{lerner2023micro} propose RHEM for coauthoring events that are labeled by the normalized number of citations of the published papers. This allows, for example, to test whether scientists are more inclined to repeat ``successful'' collaborations (assuming that success can somehow be measured by citations). Given signed hyperevents (that is, events labeled as positive or negative) one could for example assess balance \cite{cartwright1956structural,lerner2016structural,kirkley2019balance} in signed temporal hypergraphs; see \cite{lerner2013modeling} for models for dyadic signed events.

\paragraph*{Temporal decay.} Most RHEM statistics discussed in this article have been defined via the prior degree $deg(I,E_{<t})$ counting the number of previous hyperevents containing $I$ as a subset. By this definition, the popularity of nodes -- or the collective popularity of node sets -- is independent of \emph{when}, or \emph{how long ago}, these past events happened. To allow that popularity may also fade out, we can introduce a temporal decay counting recent past events more strongly than prior events that happened in the distant past. A common way to define this is via an exponential decay in which the \emph{recent prior degree} is defined by
\[
rec.deg(I,E_{<t})=\sum_{(t_m,I_m)\in E_{<t}\colon I\subseteq I_m}\exp\left((t_m-t)\frac{\log 2}{T_{1/2}} \right)\enspace,
\]
for a given half life period $T_{1/2}>0$ \cite{brandes2009networks,lerner2013modeling,schecter2021power,lerner2023relational}. The so-defined recent degree can then be used instead of the non-decaying degree in definitions of RHEM statistics. An alternative to exponential decay is a sharp dropout of past events, once the time difference exceeds a given threshold \cite{schecter2021power}. Instead of fixing the shape and speed of a decay function, one may also seek to estimate it from the data~\cite{arena2024bayesian,arena2025weighting}.

\paragraph*{Simultaneous events.} In the model definition above we assumed that event times are strictly ordered. However, in many real-world hypergraphs (e.\,g., emails within an organization or coauthoring), several or many events may have the same time stamps. Conveniently, the theory of time-to-event models provides efficient ways to handle ``tied event times'' \cite{kalbfleisch1973marginal,breslow1974covariance,efron1977efficiency,hertz1997validity} and statistical software typically can deal with multiple events occurring at the same point in time. More important is to understand the underlying assumptions. In the case of simultaneous events, RHEM assume that events with the same time stamp occur conditionally independent of each other, given the history of earlier events. The validity of this assumption depends on the given application area and also on the granularity of the time variable. To assess the robustness of findings, one may create several strictly ordered event sequences in which simultaneous events are randomly permuted and then check whether findings are affected in a meaningful way. 

\paragraph*{Time-varying risk set.} In the analysis of the \emph{Les Mis\'erables} data we assumed that each of the 80 actors could appear in any of the 288 chapters. The story told in the novel suggests that this is not true for some actors, for example, \emph{Fantine} sadly dies at the end of the first volume. We ignored this issue since we can also remark that the novel is not always chronological, so that the writer could have decided to let any actor appear in any chapter -- if necessary by jumping back or forth in time. In other applications we might want to consider more carefully who can participate in events at which point in time. This can easily be achieved by replacing the \emph{risk set} $\binom{\mathcal{I}}{|I_m|}$ in Eq.~(\ref{eq:prob_next}) by a time-varying risk set $\binom{\mathcal{I}_{t_m}}{|I_m|}$, where $\mathcal{I}_{t}$ are those nodes that could participate in events at time $t$. For example, \cite{lerner2021dynamic} analyze meeting events of former British Prime Minister Margaret Thatcher, where the possible meeting participants are her cabinet ministers at the respective time. As another example, \cite{bright2024examining,bright2024offence} apply RHEM to model co-offending networks and added actors to the risk set at their 18th birthday, since underage offenders where excluded from the data.

\paragraph*{Further model variations.}
The relation of RHEM with established models for survival analysis, or time-to-event analysis, gives access to a plethora of further model variations, including models with time-varying effects \cite{boschi2023smooth,juozaitiene2023analysing}, non-linear effects \cite{filippi2024stochastic,filippi2024modeling}, or random effects \cite{uzaheta2023random,boschi2023smooth}, model regularization \cite{friedman2010regularization,karimova2023separating}, or models embedding nodes into a latent space \cite{chen2023identifying,mulder2024latent,lakdawala2025not}. Such model variations may relax assumptions that are not necessarily valid in practice (e.\,g., homogeneity over time or nodes, or assumed linear relations between explanatory variables and response variables) and/or may increase predictive performance of the models (e.\,g., through a combination of node embeddings and model regularization).

%------------------------------------
\section{\label{sec:conclusion}Conclusion}
In this paper we reviewed relational hyperevent models (RHEM) \cite{lerner2021dynamic,lerner2023relational} and illustrated their use to define tailored null distributions for temporal hypergraphs and to test patterns such as preferential attachment, subset repetition, triadic closure, assortative mixing, and homophily. We argue that RHEM provide a versatile model family that can simultaneously reproduce in expectation observed distributions of any given vector of temporal hyperedge statistics. We sketched that RHEM have already been applied to diverse empirical settings and have been adapted to several generalizations of hypergraphs, among others to directed, typed, weighted, and signed hyperedges and to multilayer (or multimode) hypergraphs. RHEM have been applied to small networks, e.\,g., \cite{lerner2022dynamic}, as well as to networks with more than a million nodes and hyperedges \cite{lerner2025relational}. The \texttt{eventnet}\footnote{\texttt{https://github.com/juergenlerner/eventnet}} software \cite{lerner2023relational} provides a free and open-source reference implementation to compute a large selection of temporal hyperedge statistics.

The RHEM family allows to include other statistics than the one discussed in this paper. For example, \cite{lee2023temporal} define 96 different \emph{temporal hypergraph motifs}. Each of these motifs could be turned into a temporal hyperedge statistic $x(I,E_{<t})$ quantifying the change in the count of the respective motif implied by adding the possible next hyperedge $I$ to the sequence of past hyperedges $E_{<t}$. Specifying RHEM with such statistics for motif counts could be used to test a tendency for or against forming the respective temporal hypergraph motifs, compared to a null model specified with any vector of other temporal hyperedge statistics. We point out that \cite{lee2023temporal} compare empirical motif counts to a null model of random hypergraphs preserving node degrees and hyperedge sizes (as in the bipartite configuration model); RHEM, on the other hand, could be used to specify more complex null models controlling for, e.\,g., repetition of higher-order subsets, assortative mixing of various order, and/or homophily with respect to given node attributes.

The main motivation for the given analysis of the \emph{Les Mis\'erables} data was to illustrate the model. However, it nevertheless gives further empirical support to the insight (compare \cite{sarker2024higher}) that when quantifying the overrepresentation or underrepresentation of given hypergraph configurations, such as filled triads, one should not only control for observed node degrees but also for the frequencies of higher-order subsets, such as dyads contained in the respective triad (compare the discussion of the JV--MA--CO triad given above in Section~\ref{sec:statistics}). 

As pointed out in the introduction, RHEM have been applied in various fields of the social sciences -- with the purpose of testing hypothetical patterns explaining higher-order interaction. On the other hand, we are not aware of any previous discussion of RHEM to define null distributions for temporal hypergraphs. We argue that RHEM provide a suitable and versatile model family also for this purpose and hope that this paper stimulates further research in this direction.

\paragraph*{Funding:} We acknowledge financial support: J.L.\ from the Deutsche Forschungsgemeinschaft (DFG) Grant number 555455503; M.-G.H.\ from 4P-CAN project, ID 101104432, HORIZON-MISS-2022-CANCER-01, HORIZON-RIA; M.P.\ from the Slovenian Research and Innovation Agency (Javna agencija za znanstvenoraziskovalno in inovacijsko dejavnost Republike Slovenije) (Grant No. P1-0403).

\paragraph*{Data availability:} For illustration we analyze in this paper the network of actor co-appearance in chapters of Victor Hugo's novel \emph{Les Mis\'erables}, compiled by Donald Knuth \cite{knuth1993stanford}. For reproducibility, we provide the preprocessed data, augmented by the ``$female$'' variable (as described in Section~\ref{sec:data_lesmis}), in the file \texttt{jean\_events.csv}, linked from the \texttt{eventnet} Webpage \footnote{\texttt{https://github.com/juergenlerner/eventnet/tree/master/data/les\_miserables}.
}

%\bibliographystyle{unsrt}
%%\bibliographystyle{iopart-num}
%\bibliography{references}

\end{document}